\documentclass[12pt]{article}
\usepackage{graphicx}% Include figure files
\usepackage{dcolumn}% Align table columns on decimal point
\usepackage{bm}% bold math

\begin{document}
%%%%%%%%%%%%%%%%
\input psfig.sty

\begin{titlepage} 
%\begin{flushright}
%       {\bf UK/00-02}  \\
% \end{flushright}
%
\begin{center}
 
{\bf {\LARGE Heavy and Light Quarks with Lattice Chiral Fermions}}
 
\vspace{1.5cm}

{\bf K.F. Liu and S.J. Dong} \\ [0.5em]
{\it Dept.\ of Physics and Astronomy, University of Kentucky,
Lexington, KY 40506}

\end{center} 
 
%\date{\today}
 
\vspace{1cm}
 
%\vskip 12pt
%\vspace{-0.2in}
 
\begin{abstract}
The feasibility of using lattice chiral fermions which are free of
$O(a)$ errors for both the heavy and light quarks is examined. The
fact that the effective quark propagators in these fermions have the
same form as that in the continuum with the quark mass being only an
additive parameter to a chirally symmetric antihermitian Dirac
operator is highlighted. This implies that there is no distinction
between the heavy and light quarks and no mass dependent tuning of the
action or operators as long as the discretization error $O(m^2 a^2)$
is negligible. Using the overlap fermion, we find that the $O(m^2
a^2)$ (and $O(ma^2)$) errors in the dispersion relations of the pseudoscalar 
and vector mesons and the renormalization of the axial-vector current and 
scalar density are small. This suggests that the applicable range of $ma$
may be extended to $\sim 0.56$ with only 5\% error, which is a factor
of $\sim 2.4$ larger than that of the improved Wilson action. We show that the
generalized Gell-Mann-Oakes-Renner relation with 
unequal masses can be utilized to determine the finite $ma$ errors 
in the renormalization of the matrix elements for 
the heavy-light decay constants and semileptonic decay constants 
of the B/D meson.

%Given that the critical slowing down for the overlap fermion is benign for the
%light quarks, the non-perturbative renormalization is relatively straightforward, 
%and there is multi-mass algorithm for calculating the quark propagator, it 
%may be possible to study the range of quarks from  $u/d$ to $b$  
%on the same anisotropic quenched lattice with a few percent systematic error.

%We prove that the generalized Gell-Mann-Oakes-Renner relation can
%be extended to the unequal mass case so that the
%calculation of the non-perturbative renormalization
%constant such as $Z_A$ and $Z_V$ for the semi-leptonic B meson decay
%is more reliably.
%Our prelimanry result on the quenched $20^4$ lattice with overlap fermions
%at $a = 0.148 fm$ shows that the there is no problem of getting
%the hyperfine splitting of the charmonium with such a coarse lattice and
%that the renormalization constant $Z_A$ has only a mild $O(a^2)$ error. 
 
\end{abstract}

\vfill
%\pacs{11.15.Ha, 12.38.Gc, 11.30.Rd, 14.65.Dw, 14.65.Fy}

\end{titlepage}

%]  % Needed for two column format
%\narrowtext
 
  Heavy-light quarkoniums such as B and D mesons are the primary 
testing ground for understanding CP violation and obtaining the CKM matrix. 
Both experiment and theory are needed to extract relevant quantities.
For example, $|V_{ub}|/|V_{cd}|$ can be determined from the semileptonic decay rate 
of B/D meson, $B/D \rightarrow \pi l \nu$. But it depends on the 
transition form factor $|f_+(E)|$ from $B/D$ to $\pi$. $|V_{tb}^* V_{td}|$ can
be extracted from the mass difference $\Delta m_B$ of the neutral $B - \overline{B}$ mesons 
where it also depends on the B parameter $B_B$ and the yet unmeasured leptonic 
decay width $f_B$. These quantities --- $|f_+(E)|$, $B_B$, and $f_B$ are 
related to hadronic matrix elements which are best determined in
lattice QCD which is a non-perturbative approach to solving QCD with controllable 
systematic errors~\cite{kro04,ber01,ryan02}.

   The major challenge for incorporating heavy quarks with mass $m_Q$ on the lattice 
is that, in the range of lattice spacing that is amenable to numerical simulation
nowadays, the condition  $m_Q a \ll 1$ is far from being satisfied for the $b$ quark. 
There are several approaches to formulating the heavy quark on the lattice. 
The APE~\cite{all97}-UKQCD~\cite{bow00} approach is to simulate with the quark mass around 
the charm with the $O(a)$ improved Wilson action and then extrapolate to the bottom
with the guide of the heavy quark effective theory (HQET). Since the functional form for 
the mass dependence is not certain in this mass range and the extrapolated point
is very far, this results in large errors~\cite{ber01,ryan02}. 
%Furthermore, it is argued~\cite{ber01} that the $O(a)$ improvement breaks down for 
%$m_Q \rightarrow \infty$ and an explicit example~\cite{ks01} suggests that it 
%happens for $m_{\overline{MS}} a > 1/4$. 
Another approach is the non-relativistic QCD (NRQCD)~\cite{tl91}. This involves an 
expansion in terms of the heavy quark mass which is considered an irrelevant 
dynamical scale. It has the advantage that it leads to faster numerical simulation 
for the heavy quark and the correction to the static limit with higher dimensional 
operators can be incorporated in perturbation. However, there are `renormalon
shadow' effects which reflect `large perturbative uncertainties in power divergent
subtractions'~\cite{ber01}. As an effective theory, it does not have a
continuum limit. Thus, the discretization error cannot be removed by extrapolating the
lattice results to $a \rightarrow 0$. The Fermilab formulation~\cite{ekm97} bridges the 
above two approaches. For small masses, it has a continuum limit. With the 
heavy quark effective theory (HQET) approach~\cite{kro00,hhi02}, both the lattice 
spacing $a$ and the inverse of the large quark mass are treated as short-distances so
that the heavy quark discretization effects are lumped into the Wilson coefficients.
It is shown in this case that the discretization errors due to the heavy quarks can be controlled 
to allow systematic reduction of  the discretization errors for all $ma$.

%But for heavy quarks
%with $m_Q a > 1$, it is not clear if it is subjected to the `renormalon shadow' 
%effect~\cite{ber01}.  
The recent relativistic approach with anisotropic lattice~\cite{kla98} with the ratio
$\xi = a_s/a_t$ between the spatial lattice spacing $a_s$ and the
temporal spacing $a_t$ chosen to be  3--5 can alleviate the large $m_Q a_t$ problem
to a degree, but it still suffers from a large $O(m_Q^2 a_t^2)$ error~\cite{hmo02,lm02}.
To control the systematics to a few percent level for the dispersion relation, the condition
$m_Q a_t < 0.2$~\cite{hmo02} must be met which is very stringent.   

    On the other side of the approaches to heavy-light calculations, the light quark
used in the heavy-light simulation so far suffers from the well known set of problems 
associated with the lack of chiral symmetry. Take the Wilson action for example; this
ultra local action breaks chiral symmetry explicitly at finite lattice
spacing in order to lift the doublers to the cut off. As a consequence, it 
induces numerous problems. The quark mass has an additive renormalization which is
gauge configuration dependent. The quark condensate is mixed with unity
which makes it harder to calculate. There is no unambiguous correspondence
between the fermion zero modes and topology. It has $O(a)$ error and
operators in different chiral sectors mix. Although the $O(a)$ error
can be removed with the improved action and the mixing of operators can
be taken into account, the procedure nevertheless requires fine tuning and 
is usually quite involved~\cite{lss96,bgl01}. The more serious problem is the existence
of exceptional configurations. Since there is no protection by
chiral symmetry, there can be zero modes even in the presence of finite
and positive quark mass on certain gauge background configurations. This is
getting more frequent when quark mass is less than $\sim 20\, {\rm MeV}$ and
it renders the region of pion mass less than $\sim 300$ MeV inaccessible.
Unfortunately, this is the region where chiral behavior such as the
chiral logs are becoming visible. Without admission to this region of low
pion mass, reliable chiral extrapolation is not feasible~\cite{cdd03}.

 With the advent of the recent lattice chiral fermions, such as the
domain wall fermion~\cite{kap92}, the overlap fermion~\cite{neu98a}, and the 
fixed-point-action fermion~\cite{hln98},
all the above mentioned problems associated with the light quarks can be overcome
in principle. In practice, it is shown in numerical simulations of the
overlap fermion that there is indeed no additive quark mass renormalization~\cite{dll00}, 
no exceptional configurations, and the current algebra such as the
Gell-Mann-Oakes-Renner relation is satisfied to high precision~\cite{dll00}.
Besides fulfilling the promise of removing the difficulties of 
the Wilson-like fermion, the overlap fermion has turned in
extra bonuses. Its critical slowing down is quite gentle all the way to
the physical pion mass~\cite{ddh02,cdd03}; the $O(a^2)$~\cite{dll00} and $O(m^2 a^2)$
~\cite{ddh02}  errors are apparently small, and it can incorporate the multi-mass 
inversion algorithm~\cite{ehn99b}. We will concentrate on the overlap fermion
in this paper.
%Since we 
%don't know how well these features are observed in the domain wall fermion or
%the fixed-point-action fermion, we shall concentrate on the overlap fermion. 

The massless overlap Dirac operator~\cite{neu98a} is
\begin{equation}  \label{neu}
D=  1 +  \gamma_5 \epsilon (H),
\end{equation}
where $\epsilon (H) = H /\sqrt{H^2}$ is the matrix sign function of H which we
take to be the Hermitian Wilson-Dirac operator, i.e.
$H = \gamma_5 (D_w(0)-1)$. Here $D_w(0)$ is the Wilson fermion operator with
$\kappa = 1/8$. 
It is shown~\cite{lus98} that under the global lattice chiral flavor non-singlet transformation 
$\delta \psi = T \gamma_5(1 - \frac{1}{2}D)\psi, \delta\bar{\psi} = \bar{\psi} 
(1 - \frac{1}{2}D)\gamma_5 T$, the fermion action $\bar{\psi} D \psi$ is invariant
since the operator $D$ satisfies the Ginsparg-Wilson relation
 $\{\gamma_5, D\} = D\gamma_5 D$~\cite{gw82}.
It can be shown that the flavor non-singlet scalar, pseudoscalar~\cite{hhh02}, vector, and
axial~\cite{ky99,hhh02} bilinears in the form $\bar{\psi}K T (1 - \frac{1}{2}D)\psi$ 
($K$ is the kernel which includes $\gamma$ matrices) transform covariantly as 
in the continuum. 
The $1 - \frac{1}{2}D$ factor is also understood as the lattice regulator which projects out 
the unphysical real eigenmodes at $\lambda = 2$.
For the massive case, the fermion action is  $\bar{\psi}D \psi
+ m a \bar{\psi}(1 - \frac{1}{2}D) \psi$. In this case, the Dirac operator 
can be written as
\begin{equation}
D(m)=  D + m a (1 - \frac{1}{2} D).
\end{equation}
Let's consider the 
path-integral formulation of Green's function with the $\psi$ field in the 
operators and interpolation fields replaced
by the lattice regulated field $\hat{\psi} = (1 - \frac{1}{2}D)\psi$.
After the Grassmann integration, this regulator factor will be associated 
with the quark propagator in the
combination $(1 - \frac{1}{2}D) D(m)^{-1}$ which can be written as
\begin{equation}  \label{propagator}
(1 - \frac{1}{2}D) D(m)^{-1}
 %&=& \frac{1}{2} (1 - V) ( 1 + V +  \frac{m a}{2}(1 - V)))^{-1} \nonumber \\
= ( D_c + m a)^{-1}.
\end{equation}
where the operator $D_c = D/(1 - \frac{1}{2}D)$ is chirally symmetric
in the continuum sense, i.e. $\{\gamma_5, D_c\} = 0$; but, unlike $D$,
it is non-local.  This $D_c$ has been derived in the massless
case~\cite{cz98,cwz98,neu98b} in association with the quark condensate
and the solution of the Ginsparg-Wilson relation. For the massive
case, it was pointed out~\cite{chiu99,cgh99} that the mass term
$m a$ should be added to the operator $D_c$ not $D$ in the quark
propagator and Eq. (3) was derived~\cite{kn99} between the $N$-flavor
low-energy effective Dirac operator $D_N^{eff}$ from the domain wall
fermion and the truncated overlap operator $D_N$ with the overlap
operator $D$ being the $N \rightarrow \infty$ and $a_5 \rightarrow 0$
limit of $D_N$.  Here, we derive Eq. (3) from combining the lattice
regulated field $\hat{\psi} = (1 - \frac{1}{2}D)\psi$ and the inverse
of $D(m)$ to form the effective quark propagator. As a result, this
effective quark propagator should be used together with local currents
and interpolation fields without the $1 - \frac{1}{2}D$ factor.
We shall highlight
the fact that the effective quark propagator $( D_c + m a)^{-1}$ has
the continuum form, i.e. the inverse propagator is the sum of an
chirally symmetric operator and a real mass
parameter. The mass in the quark propagator is the same bare mass $m$
introduced in the fermion action. It makes no distinction between a
light quark and a heavy one, just as in the continuum.

\begin{figure}[tbh]  \label{stero}
\rotatebox{270}{
\includegraphics{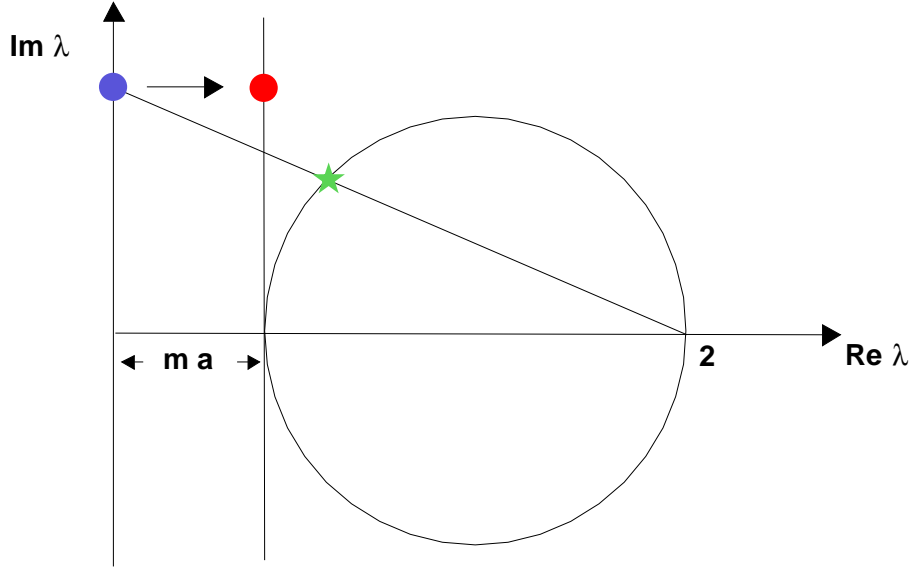}
}
\vspace*{9cm}
\caption{Stereographic projection of the eigenvalues of $D(m)$ on the circle to
the imaginary axis which is shifted by $ma$.}
\end{figure}

\begin{figure}[t]
\vspace*{7cm}
\includegraphics{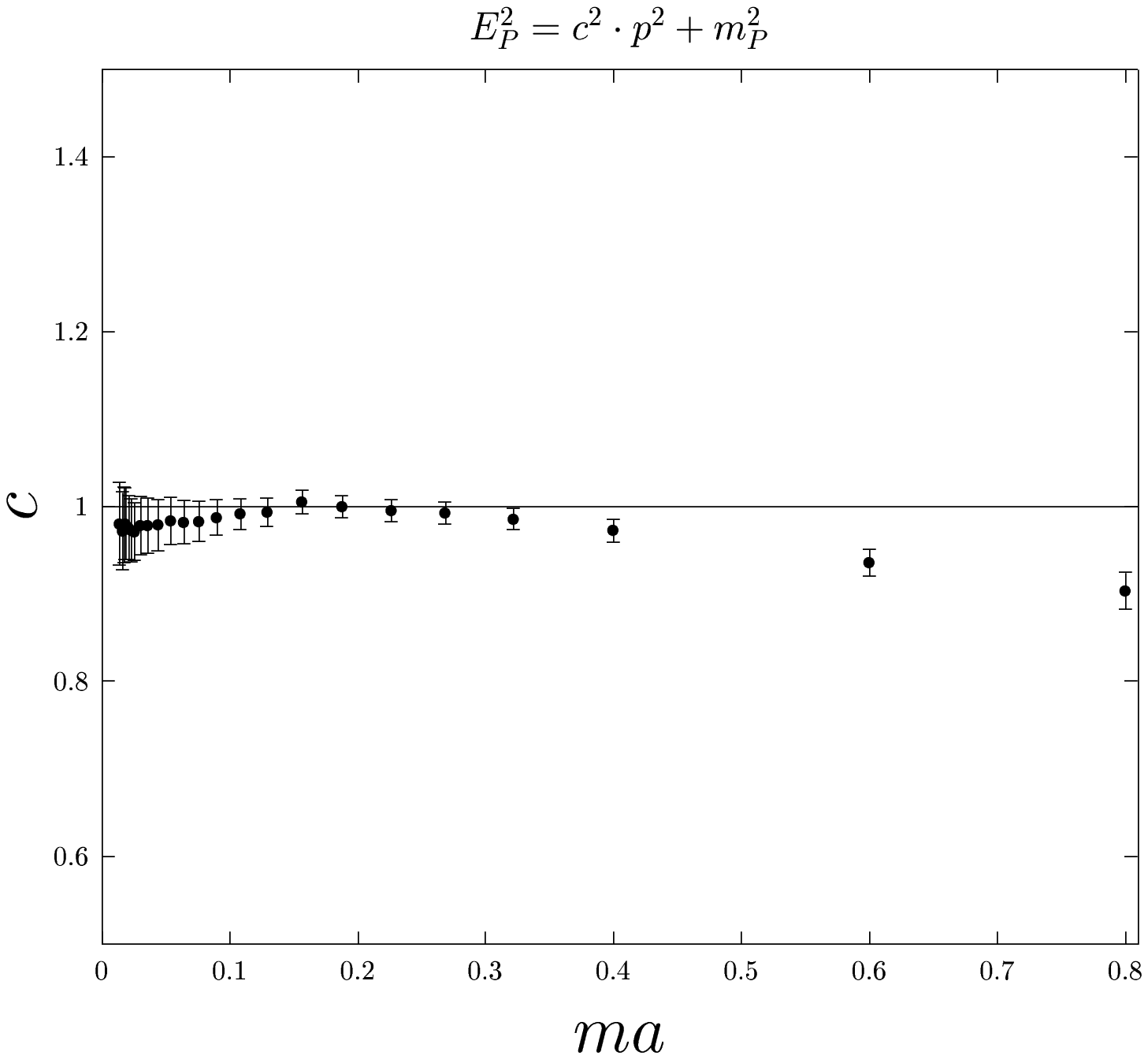}
\caption{The effective speed of light $c$ from the pseudoscalar meson dispersion relation 
as a function of $ma$.}
\label{disp_pi}
\end{figure}

It is interesting to point out that the original overlap operator $D(m)$ has 
eigenvalues lying on the circle due to the fact that $D$ satisfies 
Ginsparg-Wilson relation and is a normal matrix. This is shown in Fig. 1 where the radius is
$1 - m a/2$. On the other hand, since $D_c$ is
$\gamma_5$ hermitian, i.e. $D_c^{\dagger} = \gamma_5 D_c \gamma_5$ and anticommutes
with $\gamma_5$, it is anti-hermitian. Its eigenvalues
are simply the stereographic projection of the circle onto the imaginary 
axis as is in the massless case~\cite{cgh99}, except in the massive case, the
eigenvalue of $D_c + ma$ is shifted by $m a$ to the right. There is a 
one-to-one correspondence between the eigenstates of $D(m)$ and $D_c + ma$,
except the `north pole' at the cut-off $\lambda = 2$ which is excluded by
the regulator projector. This renders the propagator exactly like the continuum 
situation where the eigenvalues of the Euclidean Dirac operator $\not\!\!D + m$ are
distributed on the
shifted imaginary axis. Since the overlap fermion is invariant under the lattice
chiral transformation, it does not mix with dimension five operators which are not
chirally invariant. Therefore, there is no $O(a)$ nor $O(ma)$ error. 
The only question is how large the $O(m^2 a^2)$ and $O(m a^2)$ systematic errors are for 
different quark mass, be it light or heavy. We should stress that the above
discussion is not limited to the overlap fermion. It also applies to other local  
lattice Dirac operators $D$ which satisfy normality, $\gamma_5$-hermiticity, and 
the Ginsparg-Wilson relation. In general, $D_c = D/ (1- \frac{1}{2}D)$ is
the chirally symmetric operator for these lattice chiral fermions.

    We first examine the $O(m^2a^2)$ and $O(m a^2)$ errors in the dispersion relation.
It is suggested that dispersion relation is one of the places where one can 
discern the $ma$ error~\cite{kla98,hmo02}. We computed the pseudoscalar and
vector meson masses and energies at several lattice momentum, i.e. 
$p_La = \sqrt{n}\, 2\pi/La$ with $n = 0,1,2,3$. The overlap quark propagators are calculated
on the $16^3 \times 28$ quenched lattice with 80 configurations generated from
Iwasaki guage action with $a = 2.00$ fm as determined from $f_{\pi}$~\cite{cdd03}. 
Following Refs.~\cite{kla98,hmo02}, we fit the energies to the dispersion relation
\begin{equation}
(E(p)a)^2 = c^2 (pa)^2 + (E(0)a)^2
\end{equation}
where $p = 2sin(p_La/2)$. The dispersion relation is so defined such that the $ma$ error 
is reflected in the deviation of $c$ (the effective speed of light) from unity.

\begin{figure}[t]
\vspace*{7cm}
\includegraphics{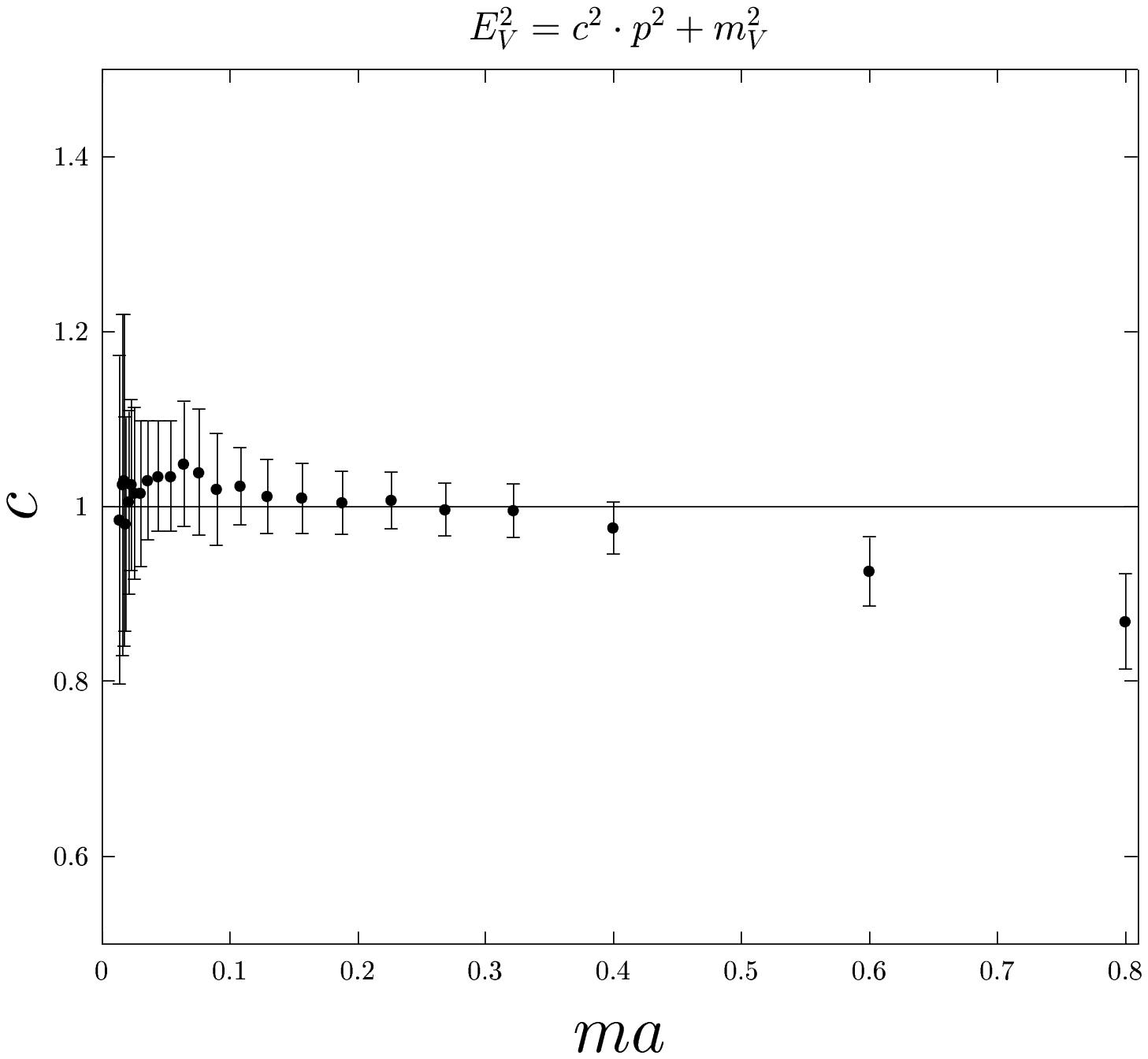}
\caption{The effective speed of light $c$ from the vector meson dispersion relation as a function
of $ma$.}
\label{disp_rho}
\end{figure}

   We see in Figs. \ref{disp_pi} and \ref{disp_rho} that the effective speed of light $c$ is 
close to unity and quite flat all the 
way to $ma \sim 0.5$. Since there is no $O(m a)$ error, we fit it with the form quadratic in $a$, 
i.e. $c = c_0 + b\, (\Lambda_{QCD}a) ma + d\, m^2 a^2$ ($\Lambda_{QCD}a = 0.188 $), and find that 
$c_0=0.982(10) , b= 0.580(346)$, and $d = -0.279(87)$ with $\chi^2/N_{dof}= 0.1$ for the
case of the pseudoscalar meson and $c_0 = 1.027(26), b= -0.32(90)$, and $ d = -0.18(22)$ with 
$\chi^2/N_{dof}= 0.6$ for the vector meson.  
Using this to gauge how large the $ma$ error is, we see that the systematic
error is less than $\sim 4\%(6\%)$ for the pseudoscalar(vector) case up to $ma \sim 0.56$. This 
$m \, a$ is $\sim 2.4$ times larger than that is admitted in the study of improved Wilson 
action~\cite{hmo02} where it is found that the $O(m^2 a^2)$ error from the anisotropy 
of the dispersion relation is less than $\sim 5\%$ when $m_Q a_t < 0.23$. 
Therefore with the overlap fermion, one can hope to extend the range of
$ma$ to $0.5 - 0.56$ where the systematic error is still reasonably small.

   Next, we address the issue of $O(ma)$ improvement. This is essential
for the Wilson-type fermions which has large $O(a)$ error. The $O(a)$ improvement
for the action is usually done with the addition of the clover term.
The operator improvement is more involved. We shall illustrate this by
considering the axial Ward identity. It has been shown~\cite{lss96} that
in the improved mass-independent renormalization scheme, the renormalized
improved axial current and pseudoscalar density have the following form from an
$O(a)$ improved action
\begin{eqnarray}  \label{ren_A}
A_{\mu}^R &=& Z_A (1 + b_A m_q a) \{A_{\mu} + c_A a \partial_{\mu}P\},
\nonumber \\
P^R &=& Z_P (1 + b_P m_q a) P,
\end{eqnarray}
where $m_q = m - m_c$ is the subtracted quark mass and $c_A, b_A$ and $b_P$
are improvement coefficients. The renormalization constants $Z_A$ and $Z_P$ are
functions of the modified coupling $\tilde{g_0}^2 = g_0^2 (1 + b_g m_q a)$.
It is argued~\cite{ber01} that as $m_Q \rightarrow \infty$ $A_4^R$ goes to 
$- \infty$ instead of the static limit, since $c_A < 0$ and 
$\partial_4 P \propto m_Q$. Even when $m_Q a \ll 1/4$, to calculate
the six parameters --- $m_c, b_A, c_A, b_p, Z_P$, and $Z_A$ non-perturbatively 
in order to satisfy the chiral Ward identity~\cite{bgl01} is
quite a task. 

\begin{figure}[t]  
\vspace*{7cm}
\includegraphics{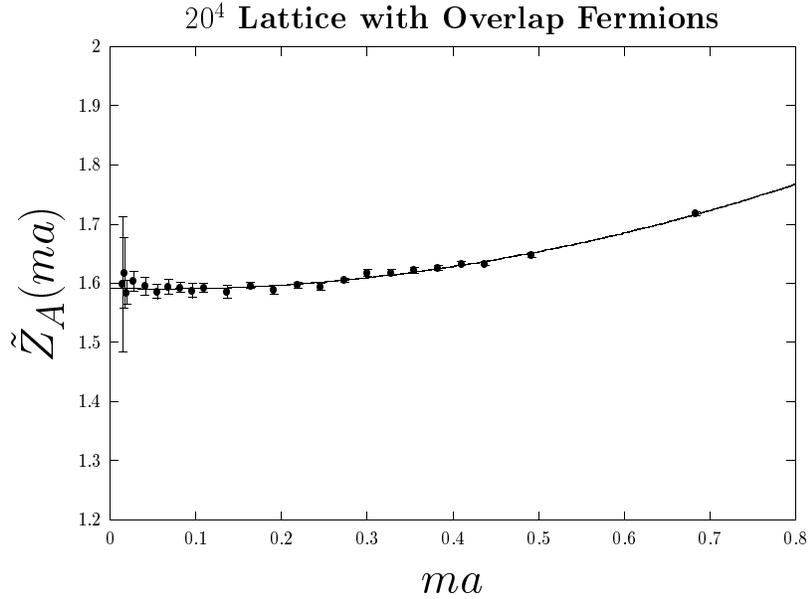}
\caption{$\tilde{Z}_A (ma)$ from the axial Ward identity on a $20^4$ lattice with $a = 0.148$ fm. 
The fitted curve which is explained in the text is plotted as the solid line.}
\label{fig:ZA}
\end{figure}

The situation with the lattice chiral fermion is much simpler.
The overlap fermion is $O(a)$ improved, and the quark mass 
is not additively renormalized which is verified numerically~\cite{dll00}. 
As a result, $b_A = c_A = b_P = m_c = 0$. 
If one uses the axial current $A_{\mu}=\bar{\psi}i\gamma_{\mu}\gamma_5\hat{\psi}$ 
for simplicity, the renormalization constant $Z_A$ can be obtained through
the axial Ward identity
\begin{equation}  \label{awi}
Z_A\partial_{\mu} A_{\mu} = 2 Z_m m Z_P P,
\end{equation}
where $P =\bar{\psi}i\gamma_5\hat{\psi}$.
Since $Z_m = Z_S^{-1}$ and $Z_S = Z_P$ due to the fact the scalar
density $\bar{\psi}\hat{\psi}$ and the pseudoscalar density $P$ are
in the same chiral multiplet, $Z_m$ and $Z_P$ cancel in Eq.~(\ref{awi})
and one can directly determine $Z_A$ non-perturbatively from the axial
Ward identity using the bare mass $m$ and bare operator $P$.
To avoid $O(a^2)$ error introduced by the derivative in Eq.~(\ref{awi}), one
can consider the axial Ward identity for the on-shell matrix elements between 
the vacuum and the zero-momentum pion state. In this case, 
\begin{equation}  \label{ZA1}
Z_A = \lim_{m \rightarrow 0,\,t \rightarrow \infty} \frac{2 m G_{PP}(\vec{p}= 0, t)}
{m_{\pi} G_{A_4P}(\vec{p}= 0, t)},
\end{equation}
where $ G_{PP}(\vec{p}= 0, t)$ and $G_{A_4P}(\vec{p}= 0, t)$ are the 
zero-momentum pseudoscalar-pseudoscalar and axial-pseudoscalar correlators.
In the mass-independent renormalization scheme~\cite{lss96}, the renormalization
factor for the axial current matrix element which takes into account the 
finite $ma$ errors can be defined from Eq. (\ref{ZA1})
without taking the massless limit and neglecting the small finite $ma$ difference 
between the renormalization of the pseudoscalar and scalar densities~\cite{zhang03,zhang04},
\begin{equation}  \label{ZAma}
\tilde{Z}_A (ma)= \lim_{t \rightarrow \infty} \frac{2 m G_{PP}(\vec{p}= 0, t)}
{m_{\pi} G_{A_4P}(\vec{p}= 0, t)},
\end{equation}
Up to $O(ma^2)$ and $O(m^2a^2)$, $\tilde{Z}_A (ma)$ is
\begin{equation}  \label{ZAtilde}
\tilde{Z}_A (ma) = Z_A (1 + b_A\, (\Lambda_{QCD}a) ma + c_A\, m^2 a^2),
\end{equation}
which is the combination of the renormalization constant $Z_A$ and the finite
$ma$ effects. The resultant $\tilde{Z}_A (ma)$
on a quenched $20^4$ lattice with overlap fermion for
$ma$ ($a = 0.148$ fm) from 0.01505 to 0.2736 were reported before~\cite{ddh02}.
Now we show the results extended to $ma = 0.684$ in Fig. \ref{fig:ZA}.
We see that, similar to the effective speed of light $c$,  it is quite flat all the 
way to $ma \sim 0.5$. In analogy to fitting $c$, we fit it with the form in
Eq. (\ref{ZAtilde}) ($\Lambda_{QCD}a = 0.188 $), and find that 
$Z_A =1.592(5) , b_A = - 0.13(9)$, and $c_A  = 0.203(22) $ with $\chi^2/N_{dof}= 0.46$.
We also have results on a $16^3 \times 28$ lattice with $a = 0.200$ fm and found an 
almost identical $ma$ behavior for $\tilde{Z}_A (ma)$~\cite{ddh03c}.
Again using this to gauge how large the $ma$ error is, we see that the systematic
error is less than $\sim 5\%$ up to $ma \sim 0.56$ which is close to the case
of $c$ for the dispersion relation for the pseudoscalar and vector mesons in 
Figs.~\ref{disp_pi} and \ref{disp_rho}.

\begin{figure}[t]
\vspace*{7cm}
\includegraphics{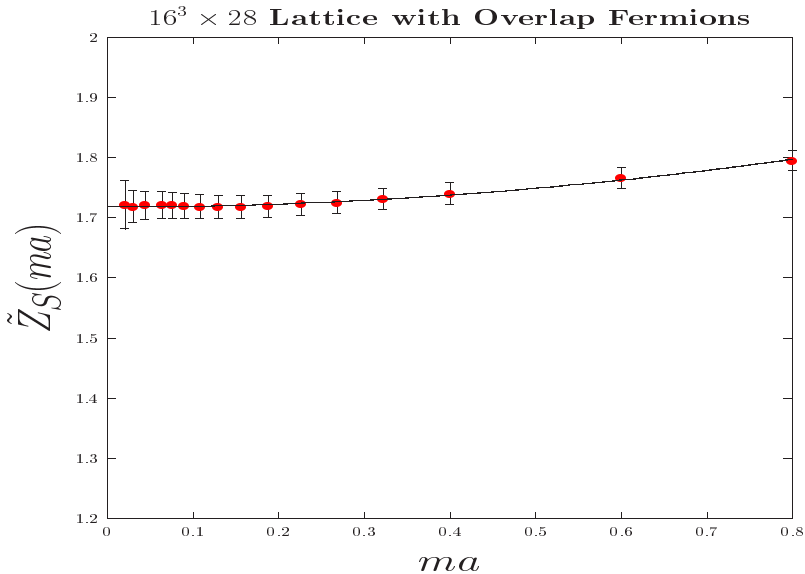}
\caption{$\tilde{Z}_S(ma)$ with $Z_S^{\overline{MS}}$(2 GeV) from the non-perturbative 
renormalization on a $16^3 \times 28$ lattice with 
$a = 0.200$ fm. The fitted curve which is explained in the text is plotted as the solid line.}
\label{Zs_2GeV}
\end{figure}

The same small discretization errors in $ma$ are reflected in other 
renormalization factors which are defined, similar to $\tilde{Z}_A (ma)$ in Eq.~(\ref{ZAtilde}), as
\begin{equation}  \label{ZGammatilde}
\tilde{Z}_{\Gamma} (ma) = Z_{\Gamma} (1 + b_{\Gamma}\, (\Lambda_{QCD}a) ma + c_{\Gamma}\, m^2 a^2),
\end{equation}
for $\Gamma = S, P, V,$ and $T$~\cite{zhang03,zhang04}. In the
mass-independent renormalization scheme~\cite{lss96}, the renormalization
constant $Z_{\Gamma}$ is a function of the gauge coupling $g_0^2$ and
the renormalization scale $\mu$, i.e. $Z_{\Gamma} = Z_{\Gamma}(g_0^2,
\mu\,a)$.  Here in Fig.~\ref{Zs_2GeV}, we show the $ma$ dependence in
$\tilde{Z}_S (ma)$ at $\mu = 2$ GeV, where the renormalization
constant $Z_S^{\overline{MS}}(2 {\rm GeV})$ is obtained from the
non-perturbative renormalization in the regularization independent
scheme~\cite{mps95,bcc02} and then perturbatively matched to the
$\overline{MS}$ scheme at the scale of 2 GeV. The results are from a
quenched $16^3 \times 28$ lattice with $a = 0.200$
fm~\cite{zhang04}. Again, we see that it is rather flat from $ma = 0$
to $ma = 0.8$. Fitting to the form in Eq. (\ref{ZGammatilde}) 
 ($\Lambda_{QCD}\,a = 0.250$) yields \mbox{$Z_S = 1.718(12),\, b_S = -0.002(194)$,} 
$c_S = 0.073(58)$. It gives a $ma$ error of 2.6\% at $ma = 0.6$.  We
should mention that similar studies for the $ma$ errors for $\tilde{Z}_A (ma)$ and
$\tilde{Z}_V (ma)$ are done with domain-wall fermions~\cite{abc02,soo03}. The $ma$
errors seem to be larger than those found here. For example, the
finite $ma$ error at $m_f a = 0.10$ is already found to be
 3 -- 4 \% which is much larger than what we obtain for
the overlap fermion at the same $ma$.

For the heavy-light quarkonium, an accurate renormalization 
for the vector and axial current is essential for the study of $f_B$, $f_D$ and
the semi-leptonic decays of the B and D mesons. Since most of the renormalization
for the composite operator with heavy and light quarks are done with perturbation 
in one loop, its $O(\alpha_s^2)$ correction can be large. In a recent calculation of $f_{B_s}$
and $f_{D_s}$ with NRQCD for the heavy quark, the $O(\alpha_s^2)$ error is estimated
to be 10\%~\cite{wdg03}. Similarly, it is pointed out in the study of $f_{D_s}$~\cite{plg03} with
fermilab heavy quark that the $O(\alpha_s)$ correction can be potentially as large as
30\%. In the following, we show a non-perturbative method which can determine the
axial heavy-light current renormalization with finite $ma$ error 
at a few percent level even with $ma$ as large as $0.5 -0.6$. This should be of help in 
determining $f_B$ and $f_D$ with much less systematic errors. 

The finite $ma$ errors in the renormalization of the matrix elements involving
$f_B$ and $f_D$ and semi-leptonic decays can be accurately determined with the
help of current algebra relations. 
The axial Ward identity for the pseudoscalar meson P decay matrix element of 
unequal masses including the finite $ma$ factor is
\begin{equation}  \label{Z_A12}
\tilde{Z}_{A}(m_1a, m_2a) \langle 0|\partial_{\mu} A_{\mu_{12}}| P\rangle 
= (\tilde{Z}_{m}(m_1a) m_1 + \tilde{Z}_{m}(m_2a) m_2)
\tilde{Z}_{P}(m_1a, m_2a) \langle 0|P_{12}| P\rangle,
\end{equation}
where $ A_{\mu_{12}} = \bar{\psi_1}i\gamma_{\mu}\gamma_5\hat{\psi_2}, 
P_{12} = \bar{\psi_1}i\gamma_5\hat{\psi_2}$ and
$\tilde{Z}_{A}(m_1a, m_2a), \tilde{Z}_{m}(ma)$, and $\tilde{Z}_{P}(m_1a, m_2a)$ are the
products of renormalization constants and their respective finite $ma$ factors.
Here  $\tilde{Z}_{P}(m_1a, m_2a)$ does not cancel out $\tilde{Z}_{m}(m_1a)/ \tilde{Z}_{m}(m_2a)$ 
except in the massless limit 
and, therefore, one cannot readily use Eq. (\ref{ZA1}) to obtain $\tilde{Z}_{A}(m_1a, m_2a)$
to account for the finite $ma$ correction. Fortunately,
one can adopt additional information from the generalized Gell-Mann-Oakes-Renner relation
for the unequal mass case, which is 
\begin{equation}   \label{GOR12}
\frac{1}{V} \int d^4 x \langle {\pi_{12}^a}^{\dagger}(x) \pi_{12}^a(0)\rangle = 
\frac{- 2 [\langle \bar{\psi_1}\hat{\psi_1}\rangle + \langle 
\bar{\psi_2}\hat{\psi_2}\rangle]}{m_1 + m_2},
\end{equation}
where $\pi_{12}^a(x) = \bar{\psi_1}\gamma_5\tau^a/2\hat{\psi_2}$.
The proof is a generalization of the equal mass case~\cite{chiu01} and it has been
proved with the staggered fermion~\cite{ks87}. In fact,
with the effective propagator in Eq.~(\ref{propagator}), a lot of the current algebra 
relations can be reproduced on the lattice with finite cutoff~\cite{cha99,ehn99b}.
From Eq.~(\ref{propagator}), we see that $\langle \bar{\psi_1}\hat{\psi_1}\rangle
= - Tr (D_c + m_1)^{-1}$ which can be written as
\begin{eqnarray}   \label{m_1}
 Tr\!\!&\!\! &\!\!(D_c + m_1)^{-1}\!\! =\!\! Tr \{(m_2\! -\! D_c)[(m_2\! -\! D_c)^{-1}
(m_1\! +\! D_c)^{-1}]\} \nonumber \\
\!\!\! &\!\! &\!\!\! = Tr  \{(m_2\! -\! D_c)[\gamma_5 (m_2\! +\! D_c)^{-1} \gamma_5 
(m_1\! +\! D_c)^{-1}]\}.
\end{eqnarray}
where we have used the property $\gamma_5 D_c \gamma_5 = - D_c$.
Similarly, one can write
\begin{eqnarray}   \label{m_2}
\!\!Tr\!\!&\!\! &\!\! (D_c + m_2)^{-1} = Tr [\gamma_5 (m_2 + D_c)^{-1}\gamma_5] \nonumber \\
\!\!\!\!\!&\!\!\!&\!\!\!\!\!= Tr  \{(m_1\! +\! D_c)[\gamma_5 (m_2\! +\! D_c)^{-1} \gamma_5 
(m_1\! +\! D_c)^{-1}]\}. 
\end{eqnarray}

\begin{figure}[t]
\vspace*{7cm}
\includegraphics{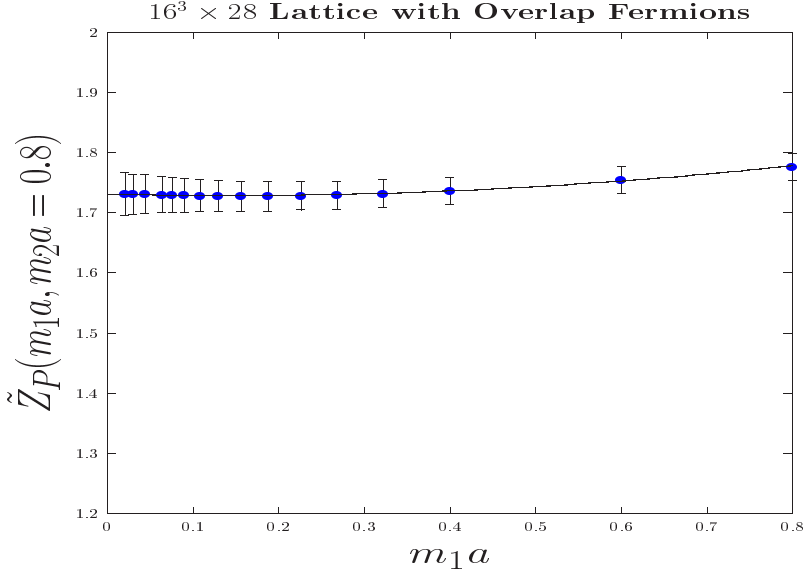}
\caption{$\tilde{Z}_{P}$ from Eq.~(\ref{Z_P12}) as a function of $m_1a$ with $m_2a$ fixed at 0.8. 
The fitted curve which is explained in the text is plotted as the solid line.}
\label{Z_P_12}
\end{figure}

Summing up Eqs. (\ref{m_1}) and (\ref{m_2}), we arrive at
\begin{eqnarray}   \label{m_12}
Tr [\gamma_5 (m_2 + D_c)^{-1} \gamma_5 (m_1 + D_c)^{-1}] \nonumber \\
\!\!\!\!= \frac{Tr[ (D_c + m_1)^{-1} + (D_c + m_2)^{-1}]}{m_1 + m_2},
\end{eqnarray}
which is just the Gell-Mann-Oakes-Renner relation for the unequal mass case in
Eq.~(\ref{GOR12}). We should note  that this relation is satisfied for
any gauge configuration, any mass, and any source for the quark propagator as is in the
equal mass case~\cite{ehn99b}. With the Gell-Mann-Oakes-Renner relation as the renormalization
condition, the same relation holds for the renormalized currents.  
%\begin{equation}   \label{GOR12}
%\frac{Z_{m_1}m_1 +Z_{m_2} m_2}{V} \int d^4 x \langle Z_{P_{12}}\pi_{12}^a(x) 
%\pi_{12}^a(0)\rangle = - 2 [Z_{S_1}\langle \bar{\psi_1}(1 - D/2)\psi_1\rangle + 
%Z_{S_2}\langle \bar{\psi_2}(1 - D/2)\psi_2\rangle].
%\end{equation}
Together, one obtains the renormalization factor which includes the renormalization constant
and the finite $ma$ correction
\begin{equation}  \label{Z_P12}
\!\!\!\tilde{Z}_{P}(m_1a, m_2a)^2 = \frac{\tilde{Z}_{S}(m_1a)\langle \bar{\psi_1}\hat{\psi_1}\rangle + 
\tilde{Z}_{S}(m_2a)\langle \bar{\psi_2}\hat{\psi_2}\rangle}{\langle \bar{\psi_1}\hat{\psi_1}\rangle 
+ \langle \bar{\psi_2}\hat{\psi_2}\rangle} \frac{ m_1 + m_2}{\tilde{Z}_{S}(m_1a)^{-1} m_1 
+ \tilde{Z}_{S}(m_2a)^{-1} m_2}.
\end{equation}
It is seen that for the massless case, the relation
$Z_P = Z_S$ is retrieved. Also, when the $O(m^2a^2)$ error
is negligible so that $\tilde{Z}_{S}(m_1a) = \tilde{Z}_{S}(m_2a)$, one finds that 
$\tilde{Z}_{P}(m_1a, m_2a) = \tilde{Z}_S(ma)= \tilde{Z}_P(ma)$. 
To assess the error for large $ma$, say $ma > 0.4$, one can first calculate the scalar
renormalization~\cite{hjl01,zhang03,zhang04} and the quark condensate to obtain $\tilde{Z}_{P}(m_1a, m_2a)$
in Eq.~(\ref{Z_P12}) which in turn determines $\tilde{Z}_{A}(m_1a, m_2a)$ from the axial Ward
identity in Eq.~(\ref{Z_A12}). This will account for the non-perturbative $ma$
error for the axial current with unequal masses. We show, in Fig.~\ref{Z_P_12}, the result of 
$\tilde{Z}_{P}(m_1a,m_2a)$ with $Z_P$ determined in the $\overline{MS}$ scheme at $2\, \rm{GeV}$ 
as a function of $m_1 a$ and with $m_2 a$ fixed at 0.8. This is obtained from  
$\tilde{Z}_S (ma)$ in Fig. 3 and the quark condensates from the equal-mass
Gell-Mann-Oakes-Renner relation. 

We see that again the $ma$ errors in $\tilde{Z}_{P}(m_1a,m_2a)$ are exceedingly small. 
Fitting it to the form $Z_{P}(1 + b_P\, (\Lambda_{QCD}a) ma + c_P\, m^2 a^2)$ ($\Lambda_{QCD}a = 0.250$)
gives $Z_{P} = 1.731(15)$, \mbox{$b_P = -0.076(245)$,} and \mbox{$c_P = 0.066(74)$.} We see that this 
value of $\tilde{Z}_{P}(m_1a, m_2a) = 1.731(15)$ for $m_1a = 0$ and $m_2 a = 0.8$ is within 
1\% of $Z_S = 1.718(12)$ (hence $Z_P$) as
we presented earlier. Through Eq.~(\ref{Z_A12}), one is expected to
obtain a non-perturbatively determined $\tilde{Z}_{A}(0, m_2a)$  which
has only  a few percent $O(\Lambda_{QCD}m a^2)$ and $O(m^2 a^2)$ errors, 
even though $m_2 a$ is as large as 0.8. Furthermore, 
a statistical error at a level of 1 -- 2\% is obtained with 80 gauge configurations. 
From this study of the renormalization of the axial current for $f_D$ and $f_B$, 
we find that even with $m_2a$ as large as $0.5 - 0.6$ the finite $ma$ error is as 
small as a few per cent. This is a good deal better than the perturbative 
determination from NRQCD or the Fermilab approach which estimates a 10\% - 30\% error
in the heavy-light decay constants~\cite{wdg03,plg03}.

%As an alternative, one can use the conserved current~\cite{ky99,hhh02} to avoid
%renormalization at the expense of a somewhat more complicated current.

   Finally, we should mention that the only major drawback of the overlap formalism is
its numerical cost which is about 50 times more than that of
the Wilson-Dirac operator at $\sim 1/5$ of the strange mass~\cite{dll00}.
This numerical overhead can be offset by extending the effective range
of $ma$ of the improved Wilson fermion by a factor of $\sim$ 2.4 (as judged on
the comparison of dispersion relations and finite $ma$ errors in the renormalization) 
and the fact that the inversion of the overlap operator accommodates multi-mass 
algorithm~\cite{ehn99b,dll00} 
in which $20 - 30$ masses can be included with only $\sim 10\%$ overhead to the
calculation of the lowest mass. For practical calculations, one may consider
an anisotropic lattice with $\xi =5$ and $a_s^{-1} = 2\, \rm{GeV}^{-1}$. Limiting
$m_Q a_t$ to 0.56, one maybe able to cover the quark spectrum from $u/d$ to $b$.
 
   To conclude, we stress that the effective quark propagator of the lattice chiral fermions 
closely parallels that of the continuum. The mass is only an additive parameter to
the chirally symmetric Dirac operator. The problems that plagued the previous light quark
formulation for lack of chiral symmetry are basically removed by the lattice chiral fermions. 
The additional desirable features of the overlap operator such as the gentle critical slowing 
down, the multi-mass inversion, and the small $O(m^2 a^2)$ and $O(m a^2)$ errors
make it suitable for the study of both light and heavy quarks without 
tuning of the actions or the operators. Whether the small $O(m^2 a^2)$ and $O(m a^2)$ errors 
hold for other quantities than the dispersion relation and the quark bilinear current renormalization 
remain to be checked. The generalized Gell-Mann-Oakes-Rener relation, 
extended to the unequal mass case, is shown to be able to facilitate the determination of the
renormalization factor $\tilde{Z}_{A}(m_1a, m_2a)$ for the calculation of the heavy-light 
decay constants and the semileptonic decay constants.
This admits the assessment of the finite $ma$ error and helps determine to which $ma$
one should carry out the calculation without large systematic errors.

This work is partially supported by DOE Grants DE-FG05-84ER40154 and DE-FG02-02ER45967.
The authors wish to thank S. Chandrasekharan, T.W. Chiu, T. Draper, I. Horv\'{a}th, 
H. Neuberger, M. L\"{u}scher, and C. Rebbi for stimulating discussions. Thanks are also 
due to J.B. Zhang for the preparation of some of the figures.

\end{document}